00 Air Quality V12

**Air quality and acute deaths in California, 2000-2012**

S. Stanley Young, CG Stat, Raleigh, NC.
Kenneth K. Lopiano, Winterville, NC
Richard L. Smith, University of North Carolina, Chapel Hill, NC

**Abstract**

Many studies have sought to determine whether there is an association between air quality and acute deaths in the US. Additionally, many consider it plausible that current levels of air quality cause acute deaths. However, several factors call causation and even association into question. Multiple testing and multiple modeling and various biases can lead to false positive findings.

Moreover, the fact that most data sets used in studies evaluating the relationships among air quality and public health outcomes are not publicly available makes reproducing the results nearly impossible. Here we have publicly available a dataset containing daily air quality levels, $PM_{2.5}$ and ozone, daily temperature levels, minimum and maximum and daily relative humidity levels for the eight most populous California air basins. Over two million death certificates were obtained from the state of California and daily death counts in the eight air basins were derived. We analyzed the dataset using a standard time series analysis, a moving median analysis, and a prediction analysis in which we use leave-one-year-out cross validation analysis to evaluate predictions. Both standard time series analysis and the moving medians analysis found little evidence for association between air quality and acute deaths. The prediction analysis process was a run as a large factorial design using different models. We use holdout predictive mean square error to assess prediction. Among the variables used to predict acute death, most of the daily death variability was explained by time of year or weather variables. In summary, neither $PM_{2.5}$ nor ozone added appreciably to the prediction of daily deaths. The empirical evidence is that current levels of air quality, ozone and $PM_{2.5}$, are not causally related to acute deaths for California.

**Introduction**

The purposes of this paper are threefold: First, we describe a data set we develop and make publically available that is useful for time-series analyses for air quality and acute deaths. Second, we provide three analyses of the data set. Third, we discuss the implications of our analysis results for the broader question of air quality and health effects.

The National Academy of Sciences, the Royal Society and the White House all support open access to data used in scientific papers **[1-3]**. In 1985 Joe Cecil **[4]** noted that "As an abstract principle, the sharing of research data is a noble goal and meets with little opposition. However, when data sharing is attempted in a particular circumstance, the conflicting interests of the parties can thwart the exchange." Our experience has been that it is difficult to get public access to data sets used in studies evaluating the relationships among public health outcomes and air quality. Consequently, we obtained raw data, built our own analysis-ready data set, and now make the data set publically available. Briefly, the data set contains daily counts of deaths in the eight most populous air basins in California for the years 2000-2012, air quality levels for ozone and $PM_{2.5}$, minimum and maximum temperature, and relative humidity.

Any large, complex observational data set, including a time series environmental epidemiology data set, can be problematic for analysis **[5-7].** In 1978, Herbert Schimmel **[5]** was one of the first to comment that claimed time series results could vary depending on how the data was analyzed. We analyze our data set using three different methods. Briefly, our first method uses moving medians with gaps, described later, to track the central distribution of the data, then computes deviations of observed values from the estimated daily central value, e.g. deviations of daily all cause deaths, ozone, $PM_{2.5}$, weather variables, etc. Our second method uses time series regression analysis **[8-9]**. Our third method examines the predictive power of air quality and weather variables of daily deaths and uses leave one year out prediction means square error**.**

The rest of this paper is organized as follows: The data sources and protocol used to generate the analysis-ready data set is described in Section 2. The methods and results of all three methods are described in Section 3. Finally, a discussion is provided in Section 4. In addition, the publicly available analysis-ready data-set is provided at http://www.unc.edu/~rls/CApollution.html and extensive supplementary material is available as part of the online publication.

**2.0 Data and Sources**

California is divided into air basins. Within air basins, weather and topography are similar. A map of California with all the air basins is given in **Figure 1.** We obtained electronic death certificate data without personal identifiers for all of California for the years 2000-2012. We chose to use the eight most populous air basins for which death counts are large enough that we could expect to see significant association if they exist.

**Figure 1:** Map of California Air Basins (Source: Webpage of the California Air Resources Board)

## 2.1 Mortality

The state of California provides access to the death public use files for the purpose of research. The available variables are given **Supplemental Death Public Use files [10]**. The cause of death is indicated by an ICD 10 code and the corresponding group cause of death is determined using the group cause of death codes from the Department of Health Services Center for Health Statistics. See California Department of Public Health, www.cdph.ca.gov for how to obtain the raw death data. The total number of deaths of individuals over 65-74 and 75+ years of age with group cause of death categorized as AllCauses or HeartLung where HeartLung deaths were attributed to "Diseases of the Circulatory System" or "Diseases of the Respiratory System". We created four outcome death categories: 65-74 AllCause, 65-74 HeartLung, 75+ AllCause, 75+ HeartLung. *Note accidental deaths were excluded from our analyses.*

## 2.2 Air Quality

The California Environmental Protection Agency's Air Resources Board provides an Air Quality Data (PST) Query tool at the following website http://www.arb.ca.gov/aqmis2/aqdselect.php **[11]**. Daily data can be retrieved for each combination of basin, day, and year. The following statistics were retrieved on July 19, 2014:

1. Daily Average $PM_{2.5}$ in $\mu g\ m^{-3}$
2. Daily Average Ozone in parts per billion (ppb)
3. Daily Max 8 Hour Overlapping Average Ozone – State in ppb
4. Daily Max 8 Hour Overlapping Average Ozone – National in ppb

## 2.3 Temperature

The Carbon Dioxide Information Analysis Center (CDIAC) maintains data from the United States Historical Climatology Network. Daily temperature data was retrieved from the following website http://cdiac.ornl.gov/ftp/ushcn_daily/ for each combination of basin, day, and year the minimum and maximum temperature were derived **[12]**.

## 2.4 Humidity

The US Environmental Protection Agency maintains daily humidity data. Daily humidity data was downloaded from http://www.epa.gov/ttn/airs/airsaqs/detaildata/downloadaqsdata.htm, **[13]**. For each combination of basin, day, and year the maximum relative humidity was derived.

## 3.0 Statistical Methods and Results

### 3.1 Robust Case Crossover Method and Results

The case crossover method is a standard method for examining acute effects in time series data [14, 15]. All individuals are cases, there are no selected controls, and time-displaced cases serve as controls. For the cases we consider a 21-day moving window. The center of the window, day 11, is the time point at issue. We compute a median of points 1-8, 14-21 and contrast that to the day 11 point. In effect we compute a deviation from the estimated time trend. The two points left and right of the time point of interest are omitted from the estimated time trend to partially remove any very local effect. The notation for our moving median is 21-5, 21 days with 5 center days removed. Our method closely resembles a bidirectional, time-stratified case crossover design [15]. The use of a 21-day window is in the range of typical time windows [15] used in air pollution studies. The omitting of points near the point of interest is used in an attempt to decrease the influence of spikes on the moving median [15]. The use of a moving median with a gap is novel so far as we know. A SAS JMP addin [16] is available from the first author to compute this time series trend estimate. This moving median is also used to estimate time trends in air quality and meteorological variables. Deviations from that trend are computed, giving a deviation for each day where the trend can be computed. A touted advantage of a case crossover design is that the cases within the window are expected to be very similar with respect to many covariates; there is time-local control.

Moving median deviations and simple two-way plots of raw data and are presented first to give the reader a sense of part of the data set. We use South Central Coast (Los Angeles) data as the number of daily deaths is high and pollution levels in LA are relatively high. If there are pronounced effects we should see them.

Partial correlations are computed by first predicting each variable using multiple linear regression of the other variables and then correlating the residuals for each pair of variables. Partial correlations are used for inferring potentially causal relationships among variables as they are adjusted for other variables under consideration. We use the JMP addin Partial Correlation Diagram.

**Temperature**
The analysis in this section is limited to the South Central Coast air basin, Los Angeles. The response variable was daily all cause deaths for people 75 and older. To illustrate the utility of the moving median analysis consider the relationship between death and maximum temperature, **Figure 2**. It is shown that as maximum temperature increases the number of daily deaths decreases. However, the effect may be confounded by other variables that are seasonal. Note the seasonal relationship between death and time of year, **Figure 3**. The moving median is used to model the relationship between death and time of year and temperature and time of year with a 21-5 moving median. Deviations from the moving medians are then calculated for both all cause deaths and maximum temperature, denoted as D All Cause 75 and D Tmax, respectively. The relationship between these two variables illustrates that a spike in temperature as measured by a large deviation from the moving median is complemented by a spike in daily deaths as measured by a deviation from the moving median, **Figure 4. Sup 3.1 A** gives a 2x2 multiple scatter plot of All Cause deaths 65-74 and 75+ by mint and maxT. Older individuals appear to be sensitive to an increase in maxT.

Figure 2. All Cause daily deaths of people 75 and older are plotted against the maximum daily temperature. Each dot represents the number of daily deaths. Accidental deaths are removed. Non-parametric density contours are added to help visualize the relationship.

Figure 3. Daily all cause deaths are plotted against time, 2000-2012. A spline is fit to the data to help visually track the density of the points.

Figure 4. Deviations of daily deaths are plotted against deviation of daily temperature. A spline indicates that as the there is a spike in daily maximum temperature there is an increase in daily deaths. Three sets of points are colored, red, green and aqua where the spike in temperature is 15-28 degrees higher than the moving median.

**Ozone**
A similar analysis was conducted to assess the relationship between all cause deaths for people 75 and older (deaths) and daily max 8 hour lag ozone (ozone). The bivariate relationship between deaths and ozone is shown in **Figure 5.** The moving medians for both time series are plotted in **Figure 6**. The result is a smoothed relationship between each of the variables with time. Note both ozone and deaths are seasonal over time and are out of phase. As ozone increases, deaths decrease. Deviations from the moving medians are calculated and the relationship between the deviations is illustrated graphically in **Figure 7.** Note there is no apparent effect of ozone on deaths. **Sup 3.1 B** gives multiple scatter graphs with lags of 0, 1, and 2 days of Deviations of Death vs Deviations of ozone from their moving medians. There is no apparent association of deaths with ozone. There are 96 scatter graphs, 8 air basins, 4 death outcomes and 3 lags.

Figure 5. All Cause daily deaths of people 75 and older are plotted against the maximum 8-our daily ozone. Each dot represents the number of daily deaths. Accidental deaths are removed. Non-parametric density contours are plotted to help overcome the overprinting.

Figure 6. The moving medians for all cause deaths and ozone are plotted against time, 13 years, red o for deaths and blue + for ozone.

Figure 7. Deviations of daily deaths are plotted against deviation of daily ozone. A spline indicates that when the there is a spike in daily ozone, there is little or no effect on daily deaths. The density does not go from lower left to upper right. Dramatic increases in ozone have essentially no effect on deaths.

**PM$_{2.5}$**

A similar analysis was used to assess the relationship between all cause deaths and PM2.5, see **Figures 8** and **9**. Again, there is no apparent effect of PM2.5 on deaths. **Sup 3.1 C** gives multiple scatter graphs with lags of 0, 1, and 2 days of Deviations of Death vs Deviations of PM$_{2.5}$ from their moving medians. There is no apparent association of deaths with PM$_{2.5}$. There are 96 scatter graphs, 8 air basins, 4 death outcomes and 3 lags.

Figure 8. All Cause daily deaths of people 75 and older are plotted against the PM$_{2.5}$. Each dot represents the number of daily deaths. Accidental deaths are removed. Non-parametric density contours are plotted to help overcome the overprinting. A spline fit to the data indicates no association of deaths with PM$_{2.5}$ levels.

Figure 9. Deviations of All Cause daily deaths of people 75 and older are plotted against the deviations of PM$_{2.5}$ from yearly medians. There is no indication of any association.

**Partial Correlations**
See **Figure 10**. There are strong positive (thick red lines) and negative (thick blue lines) partial correlations among the weather variables. PM$_{2.5}$ and ozone have high partial correlations with MAXRH and Tmax. Tmin has a high partial correlation with MAXRH and Tmax. All Cause mortality has weak, positive correlations with Tmin and Tmax. Partial correlations between All Cause mortality and air pollution variables are inconsequential (thin lines and multiple testing adjusted p-values, results not shown).

Figure 10. Partial correlation diagram among the moving median deviations. Partial correlations between All Cause mortality and air pollution variables are inconsequential.

**3.2 Time Series Analysis of California Mortality Data**
The time series model is adapted from models previously used for the National Morbidity, Mortality and Air Pollution (NMMAPS) data series; see in particular **[8, 9, 18].** The code used for the results in the present paper is at www.unc.edu/~rls/EpiTimeSeriesCodeRLS.txt **[19]**.

The basic model is of the form

$$\text{Log}(\mu_t) = \text{Overall Mean} + \text{DLM}(l_1,\ldots,l_k) + s(t, n_{yr}*df_0) + \text{DOW}$$
$$+ s(M_1(t); df_1) + s((M_1(t-1) + M_1(t-2) + M_1(t-3))/3; df_2)$$
$$+ \ldots + s(M_p(t); df_1) + s((M_p(t-1) + M_p(t-2) + M_p(t-3))/3; df_2) \qquad (1)$$

where

> $\mu_t$ is expected number of deaths on day t;
>
> DLM($l_1,\ldots,l_k$) refers to a (linear) distributed lag model for the air pollution variable; this includes regression terms $\beta_1(X(t-l_1) + X(t-l_2) + \ldots + X(t-l_k))/k + \beta_2\{ X(t-l_2) - (X(t-l_1) + X(t-l_2) + \ldots + X(t-l_k))/k\} + \ldots + \beta_k\{ X(t-l_k) - (X(t-l_1) + X(t-l_2) + \ldots + X(t-l_k))/k\}$ where the lead coefficient $\beta_1$ represents the mean rise in mortality per one unit rise in air pollution X, distributed over lags $l_1,\ldots,l_k$;
>
> s(t, $n_{yr}$*$df_0$) refers to a natural spline on time variable t over $n_{yr}$ years with $df_0$ degrees of freedom per year; this represents the long-term trend (including seasonal component);
>
> DOW is a day of week component (treated as a factor variable with 6 degrees of freedom);
>
> s($M_1$(t); $df_1$) represents a nonlinear trend on current-day value of met variable $M_1$ with $df_1$ degrees of freedom;
>
> s(($M_1$(t-1) + $M_1$(t-2) + $M_1$(t-3))/3; $df_2$) represents a nonlinear trend on the average of the three previous days of met variable $M_1$ with $df_2$ degrees of freedom;
>
> other met variables $M_2,\ldots,M_p$ are treated similarly to $M_1$;
>
> days with missing data are omitted from the analysis;
>
> the model is fitted as a generalized linear model with log link and *quasipoisson* mean-variance structure; this is similar to assuming a Poisson distribution but with an additional parameter representing overdispersion.

In addition to fitting model (1) with all the variables, we have also fitted the model without the air pollution component and dropping some of the meteorological terms. A likelihood ratio test is conducted when each of the terms from s($M_1$(t); $df_1$) to s(($M_p$(t-1) + $M_p$(t-2) + $M_p$(t-3))/3; $df_2$) is dropped; this is an additional check that the selection of meteorological terms is appropriate.

For the California analysis, we have used three meteorological variables: daily maximum temperature, daily minimum temperature and daily mean relative humidity. Previous NMMAPS studies including **[8]** have used two meteorological variables, daily mean temperature and dewpoint, but otherwise the same model form as above. For the degrees of freedom, in previous studies $df_0$ has been typically taken between 7 and 12, $df_1$ and $df_2$ between 3 and 6; we have varied these by trial and error to understand the sensitivity of the analysis to these choices.

The most critical component of the model (1) is the selection of lags $l_1,\ldots,l_k$ to represent the air pollution component. The NMMAPS analyses in Bell **[15]** and Smith **[8]** used lags 0, 1, 2,…,6 (in some cases with an additional refinement, the *constrained distributed lag* model in which some of the coefficients $\beta_2,\ldots,\beta_k$ are constrained to be equal; however, this usually has only minor impact on the important coefficient $\beta_1$). Other common approaches use any of lags 0, 1, 2 in a single-lag model, or averages over any combination of lags 0, 1, 2, 3. For the present study, we have tried different combinations of lags to look for the lag combination that best represents the air pollution effect. We believe this approach to be justified in view of the weak evidence for any air pollution effect in these dataset; however, in view of the selection bias inherent in such an approach, we caution against over-interpretation of such results, especially in cases where the p-value is over 0.01 or the result highly depend on the selection of a particular combination of lags.

**South Coast Air Basin**

The approach outlined in the previous section is applied to data from each of eight California air basins, **Figure. 1.** Because they are the two most populated air basins, we concentrate initially on the South Coast air basin (which includes Los Angeles, Orange, Riverside and San Bernardino Counties) and the San Francisco Bay air basin (San Francisco, Marin, Sonoma, Napa, Solano, Contra Costa, Alameda, Santa Clara and San Mateo counties). For the response variable in this analysis, we use total non-accidental mortality among people aged 65 and over.

Fitting the meteorological model alone, in **Table 1** we tabulate the p-value associated with dropping each of the six terms in turn. Five of the meteorological variables are very highly significant; the only exception is current-day relative humidity. This result is based on the particular choices $df_0=7$, $df_1=df_2=6$, but the overall conclusion is robust against alternative values of those three degree of freedom parameters.

| Variable | Lags | p-value |
|---|---|---|
|  |  |  |
| Daily Max Temperature | Current day 0 | <1 e-16 |
| Daily Max Temperature | Mean of 1,2,3 | 4.6 e-7 |
| Daily Min Temperature | Current day 0 | 2.5 e-4 |
| Daily Min Temperature | Mean of 1,2,3 | 2.4 e-5 |
| Mean Daily Relative Humidity | Current day 0 | 0.18 |
| Mean Daily Relative Humidity | Mean of 1,2,3 | 1.5 e-10 |

**Table. 1:** Statistical significance of meteorological components: based on model (1) without air pollution component and with $df_0=7$, $df_1=df_2=6$, fitted to nonaccidental mortality for ages 65 and up, South Coast air basin.

In subsequent analyses, we have retained all six meteorology components; this is to ensure consistency across different air basins and to avoid the analysis being biased by overuse of statistical significance tests; however, **Table. 1** is evidence that we have identified appropriate meteorological variables for the overall analysis.

We now consider addition air pollution variables to the meteorological model in **Table 1**. Initially, we concentrate on ozone. **Table 2** shows the coefficient estimates, standard error (SE), t-value and p-value associated with ozone at various combination of lags. The units here are percent rise in mortality per 10 ppb rise in ozone. The strongest positive coefficient is based on lags 0, 1, 2 and 3, for which the model predicts a 0.1% rise in mortality per 10 ppb rise in ozone. However, neither this nor any of the other values in the table comes anywhere close to being statistically significant. This is for 13 years of data over one of the most densely populated areas of the US – if there is an ozone-mortality effect in California, we ought to see it here.

| Lags Included | Estimate | SE | t-value | p-value |
|---|---|---|---|---|
| | | | | |
| 0 | 0.0870 | 0.1135 | 0.77 | 0.44 |
| 1 | -0.0472 | 0.1136 | -0.42 | 0.68 |
| 2 | 0.0471 | 0.1141 | 0.41 | 0.68 |
| 0,1 | 0.0266 | 0.1315 | 0.20 | 0.84 |
| 1,2 | 0.0002 | 0.1330 | 0.00 | 1.00 |
| 0,1,2 | 0.0825 | 0.1507 | 0.55 | 0.58 |
| 0,1,2,3 | 0.1222 | 0.1673 | 0.73 | 0.46 |
| 0,1,2,3,4 | 0.0941 | 0.1802 | 0.52 | 0.60 |
| 0,1,2,3,4,5 | 0.0096 | 0.1905 | 0.05 | 0.96 |
| 0,1,2,3,4,5,6 | -0.0479 | 0.1992 | -0.24 | 0.81 |

**Table. 2:** Statistical significance of ozone component with various combinations of lags: based on model (1) $df_0=7$, $df_1=df_2=6$. Estimate is percent rise in mortality for 10 ppb rise in ozone. South Coast air basin; response variable is non-accidental mortality aged 65 and over.

The same analysis was tried using $PM_{2.5}$ in place of ozone, with results shown in **Table 3.** In this case, several of the estimates appear to be statistically significant with a p-value <0.05 (smallest value 0.017), but all the statistically significant values are negative, which is not biologically plausible. We conclude that either the small p-values are an artifact of the selection effect already mentioned, or there is some other biological mechanism, such as confounding by some other pollutant, that explains these results.

| Lags Included | Estimate | SE | t-value | p-value |
|---|---|---|---|---|
| 0 | 0.1261 | 0.0998 | 1.26 | 0.21 |
| 1 | -0.1966 | 0.0990 | -1.99 | 0.05 |
| 2 | -0.2121 | 0.0995 | -2.13 | 0.03 |
| 0,1 | -0.0425 | 0.1144 | -0.37 | 0.71 |
| 1,2 | -0.2720 | 0.1151 | -2.36 | 0.018 |
| 0,1,2 | -0.1133 | 0.1294 | -0.88 | 0.38 |
| 0,1,2,3 | -0.1636 | 0.1409 | -1.16 | 0.25 |
| 0,1,2,3,4 | -0.1611 | 0.1499 | -1.07 | 0.28 |
| 0,1,2,3,4,5 | -0.2609 | 0.1582 | -1.65 | 0.10 |
| 0,1,2,3,4,5,6 | -0.2435 | 0.1659 | -1.47 | 0.14 |

**Table 3:** Statistical significance of $PM_{2.5}$ component with various combinations of lags: based on model (1) $df_0=7$, $df_1=df_2=6$. Estimate is percent rise in mortality for 10 μg/m$^3$ rise in $PM_{2.5}$. South Coast air basin; response variable is non-accidental mortality aged 65 and over.

In these analyses, the overdispersion parameter was of the order of 1.07 – in other words, the variance of the mortality variables is inflated by a factor of 1.07 compared with the Poisson distribution. This is typical for this kind of analysis and does not indicate a problem. A much larger overdispersion parameter could indicate some important missing covariates.

**San Francisco Bay Air Basin**

So far, we have only considered one air basin. The second most populated is San Francisco Bay, which has substantially different weather patterns and demographics from the Los Angeles area. Therefore, the entire analysis has been repeated for this air basin, as a test of how robust the analyses are for different regions of the state.

**Table 4** shows the statistical significance of the individual meteorology components, analogous to **Table 1** for the South Coast air basin. The main difference from **Table 1** is that neither of the components due to relative humidity is statistically significant. (Although not reported in the table, if both relative humidity components – current day and the average of lags 1, 2, 3 – are dropped together, rather than one at a time, we also do not get a statistically significant component due to relative humidity.) In the following analyses, to maintain consistency of analysis methods across different air basins, the main results are still reported including relative humidity, but to assess the sensitivity to this component, some of the analyses have been repeated omitting relative humidity altogether.

| Variable | Lags | p-value |
|---|---|---|
|  |  |  |
| Daily Max Temperature | Current day 0 | 9.05E-11 |
| Daily Max Temperature | Mean of 1,2,3 | 0.0071 |
| Daily Min Temperature | Current day 0 | 0.0019 |
| Daily Min Temperature | Mean of 1,2,3 | 0.043 |
| Mean Daily Relative Humidity | Current day 0 | 0.41 |
| Mean Daily Relative Humidity | Mean of 1,2,3 | 0.32 |

**Table 4:** Statistical significance of meteorological components: based on model (1) without air pollution component and with $df_0=7$, $df_1=df_2=6$, fitted to nonaccidental mortality for ages 65 and up, San Francisco Bay air basin.

**Table 5** shows the results when ozone is added to the analysis. As with our earlier analyses for the South Coast air basin, none of the estimates of the ozone effect at various lags is statistically significant at the 0.05 level. However, two of the analyses (with lag 0 alone, and with lags 0 and 1 together) are statistically significant with a p-value of about .02 if the relative humidity component is omitted. This result illustrates the principle that if enough different models are tried, it is usually possible to find some model that gives a statistically significant result: it does not imply that the result is significant in any practical sense. It should also be noted, however, that all the coefficients of models that include lag 0 are similar in magnitude (between 0.3 and 0.6): the variation in p-values is mostly due to their standard errors.

| Lags Included | RH included? | Estimate | SE | t-value | p-value |
|---|---|---|---|---|---|
| | | | | | |
| 0 | yes | 0.4464 | 0.2471 | 1.81 | 0.071 |
| 1 | yes | 0.1889 | 0.2413 | 0.78 | 0.43 |
| 2 | yes | -0.1560 | 0.2442 | -0.4 | 0.52 |
| 0,1 | yes | 0.4909 | 0.3030 | 1.62 | 0.11 |
| 1,2 | Yes | 0.0225 | 0.2947 | 0.08 | 0.94 |
| 0,1,2 | Yes | 0.3281 | 0.3502 | 0.94 | 0.35 |
| 0,1,2,3 | Yes | 0.4210 | 0.3927 | 1.07 | 0.28 |
| 0,1,2,3,4 | Yes | 0.4716 | 0.4167 | 1.13 | 0.26 |
| 0,1,2,3,4,5 | Yes | 0.4703 | 0.4310 | 1.09 | 0.28 |
| 0,1,2,3,4,5,6 | Yes | 0.3325 | 0.4448 | 0.75 | 0.45 |
| 0 | No | 0.4838 | 0.2121 | 2.28 | 0.023 |
| 0,1 | No | 0.5948 | 0.2604 | 2.28 | 0.022 |

**Table 5:** Statistical significance of ozone component with various combinations of lags: based on model (1) $df_0=7$, $df_1=df_2=6$. Relative humidity is omitted from some of the analyses. Estimate is percent rise in mortality for 10 ppb rise in ozone. San Francisco Bay air basin; response variable is non-accidental mortality aged 65 and over.

**Table 6** shows the corresponding results for $PM_{2.5}$, where again relative humidity has been omitted from some of the analyses to illustrate the sensitivity to this component. Our conclusions are similar: some rows of this table show a statistically significant effect with a p-value of the order 0.02, but taking account of the number of models examined in order to achieve this result, it is unlikely to be of practical significance.

The overdispersion parameter for these analyses was around 1.05.

| Lags Included | RH included? | Estimate | SE | t-value | p-value |
|---|---|---|---|---|---|
| 0 | Yes | 0.3031 | 0.2362 | 1.28 | 0.20 |
| 1 | Yes | 0.1235 | 0.2373 | 0.52 | 0.60 |
| 2 | Yes | 0.3769 | 0.2312 | 1.63 | 0.10 |
| 0,1 | Yes | 0.3968 | 0.2700 | 1.47 | 0.14 |
| 1,2 | Yes | 0.4614 | 0.2679 | 1.72 | 0.09 |
| 0,1,2 | Yes | 0.5903 | 0.3067 | 1.92 | 0.05 |
| 0,1,2,3 | Yes | 0.5688 | 0.3297 | 1.72 | 0.08 |
| 0,1,2,3,4 | Yes | 0.5042 | 0.3482 | 1.45 | 0.15 |
| 0,1,2,3,4,5 | Yes | 0.5500 | 0.3634 | 1.51 | 0.13 |
| 0,1,2,3,4,5,6 | Yes | 0.4884 | 0.3767 | 1.30 | 0.19 |
| 0,1,2,3 | No | 0.5712 | 0.3123 | 1.83 | 0.07 |
| 0,1,2,3,4 | No | 0.6518 | 0.3341 | 1.95 | 0.05 |
| 0,1,2,3,4,5 | No | 0.8169 | 0.3535 | 2.31 | 0.021 |
| 0,1,2,3,4,5,6 | No | 0.7737 | 0.3702 | 2.09 | 0.037 |

**Table 6:** Statistical significance of $PM_{2.5}$ component with various combinations of lags: based on model (1) $df_0=7$, $df_1=df_2=6$. Relative humidity is omitted from some of the analyses. Estimate is percent rise in mortality for 10 µg/m³ rise in $PM_{2.5}$. San Francisco Bay air basin; response variable is non-accidental mortality aged 65 and over.

**Combining Results Across Air Basins**

In the NMMAPS papers on ozone **[9, 18]**, the single-city analyses were repeated for up to 98 US cities for which ozone and mortality data were available. They were then combined across cities using a hierarchical model analysis, based on an algorithm originally due to Everson and Morris **[21]** and coded by Roger Peng into the R function "tlnise" **[22]**. The same method is used here to produce estimates that are combined across all eight air basins in our study. It would not be practicable (or interpretable) to repeat all the analyses for every combination of meteorological variables, lags of the pollutant variable, or degrees of freedom for the spline components of the model. Therefore, some choices were made, guided by the analyses already conducted for the South Coast and San Francisco Bay air basins, as follows:

1. All analyses used all six meteorological variables.
2. The degree of freedom parameters were set to be respectively 7, 6 and 6, for $df_0$, $df_1$ and $df_2$.
3. For both ozone and $PM_{2.5}$, only certain combinations of lags were tried.

The results of this analysis are shown in **Table 7.** None of the analyses show a statistically significant effect when combined across all eight air basins.

| Variable | Lags | Estimate | SE | t-value | p-value |
|---|---|---|---|---|---|
|  |  |  |  |  |  |
| Ozone | 0,1 | 0.3376 | 0.2434 | 1.39 | 0.17 |
| Ozone | 0,1,2 | 0.3165 | 0.2466 | 1.28 | 0.20 |
| Ozone | 0,1,2,3 | 0.4149 | 0.3260 | 1.28 | 0.20 |
| PM2.5 | 0,1 | 0.0126 | 0.2034 | 0.06 | 0.95 |
| PM2.5 | 0,1,2,3 | -0.0006 | 0.2464 | 0.00 | 1.00 |
| PM2.5 | 0,1,2,3,4,5 | 0.0689 | 0.2799 | 0.25 | 0.81 |

**Table 7:** Combined results across all eight air basins.

All the analyses in this paper so far are based on total non-accidental mortality for ages 65 and up. The analysis was repeated using (a) total non-accidental mortality for all ages, (b) respiratory deaths aged 65 and up, (c) circulatory deaths aged 65 and up, (b) combined respiratory and circulatory deaths aged 65 and up. None of these produced a statistically significant result in the combined analyses.

The results of **Table 7** were also repeated with the choices $df_0=7$, $df_1=6$, $df_2=6$ replaced by (a) $df_0=10$, $df_1=6$, $df_2=6$, (b) $df_0=7$, $df_1=3$, $df_2=3$, (c) $df_0=10$, $df_1=3$, $df_2=3$. The analysis of **Table 7** was also repeated with relative humidity omitted from the analysis. None of these changes produced a statistically significant result in any of the combined analyses.

**[Point to Sup 3.2 A and Sup 3.2 B here]**

**Comparisons with NMMAPS**

We have pointed out that the statistical methods of this paper are similar to those of the NMMAPS study; see in particular **[8, 15]**, but they are not identical. Those papers also included an interaction effect between age and long-term trend, and the meteorological variables were daily mean temperature and dewpoint, rather than those of the present paper. What happens if we use exactly the same methods for the two datasets?

To investigate this question, we recompiled the NMMAPS dataset but using tmax, tmin and daily max relative humidity as the meteorological variables. (Those variables are all in the NMMAPS dataset, but were not used in the previously cited papers.) The dataset was analyzed using the same computer code as the other analyses in this paper, applied to deaths aged 65 and over analyzed as a single age group (no interactions). We took $df_0=7$, $df_1=df_2=6$ as in most of the analyses in this paper, and the distributed lag structure based on lags 0 through 6.

Since the rest of this paper is concerned with California data, we concentrated on the California cities in the NMMAPS database. **Table 8** shows results for each city, and the combined result for all 12 California cities. Also shown in **Table 8** is the national result, in which the 12 California cities were combined with 86 other US cities, reanalyzed using the software of the present paper.

| City | Estimate | SE | t-value | p-value |
|---|---|---|---|---|
| Bakersfield | 0.7031 | 0.9970 | 0.71 | 0.48 |
| Fresno | 0.1577 | 0.9520 | 0.17 | 0.87 |
| Los Angeles | 0.1941 | 0.2199 | 0.88 | 0.38 |
| Modesto | 0.3027 | 1.5057 | 0.20 | 0.84 |
| Oakland | 0.8943 | 1.0210 | 0.88 | 0.38 |
| Riverside | 0.0255 | 0.6019 | 0.04 | 0.97 |
| Sacramento | -0.0913 | 0.8334 | -0.11 | 0.91 |
| San Bernardino | 0.7358 | 0.6330 | 1.16 | 0.25 |
| San Diego | 0.1080 | 0.4717 | 0.23 | 0.82 |
| San Jose | -0.0481 | 0.9756 | -0.05 | 0.96 |
| Santa Ana Anaheim | 0.1231 | 0.4815 | 0.26 | 0.80 |
| Stockton | 0.9981 | 1.3775 | 0.72 | 0.47 |
| All CA | 0.2485 | 0.2307 | 1.08 | 0.28 |
| National | 0.2873 | 0.0915 | 3.14 | 0.0017 |

**Table 8:** Estimates for the ozone effect in 12 California cities from the NMMAPS study (San Francisco omitted because of lack of ozone data). Also shown are the combined results from all 12 cities under "All CA", and the combined results of all 98 US cities included in the NMMAPS ozone study. Applied to all deaths aged 65 and up, using tmax, tmin and maximum relative humidity as the three meteorological variables, and a distributed lag model for ozone covering lags 0-6.

The last result shows a combined estimate of 0.287 (percent rise in mortality per 10 ppb rise in 8-hour daily max ozone) and a standard error (more precisely, posterior standard deviation) of 0.0915. By comparison, the result quoted in Smith **[9]** was a combined estimate of 0.411 and a posterior standard deviation of 0.080. Just to make a further comparison with the results of Smith **[9],** the method of the present paper was repeated with mortality data from all age groups 55 and up (the same as in the original NMMAPS analyses) – in this case our estimated combined national coefficient, using the meteorological model of the present paper, rises only very slightly, from 0.287 to 0.300. Therefore, the difference in combined estimates compared with Smith **[9]** appears to be due to the different meteorological variables used and not to the different treatments of age groups. It seems plausible that the treatment of meteorology in the present paper (in particular, the separate use of tmax and tmin) is superior to the treatment in the earlier NMMAPS papers, resulting in a lower estimate of the ozone effect because of less confounding by weather.

**Nonlinear Distributed Lag Models**

In this section, we consider an extension of the preceding analyses that allows for the leading air pollution term to be nonlinear.

Specifically, where we have previously defined the DLM($l_1,\ldots,l_k$) to have components

$$\beta_1(X(t-l_1) + X(t-l_2) + \ldots + X(t-l_k))/k + \beta_2\{X(t-l_2) - (X(t-l_1) + X(t-l_2) + \ldots + X(t-l_k))/k\} + \ldots + \beta_k\{X(t-l_k) - (X(t-l_1) + X(t-l_2) + \ldots + X(t-l_k))/k\}, \quad (2)$$

we now replace the term $\beta_1(X(t-l_1) + X(t-l_2) + \ldots + X(t-l_k))/k$ by a nonlinear term of form

$$s((X(t-l_1) + X(t-l_2) + \ldots + X(t-l_k))/k; df_3), \quad (3)$$

in other words, a nonlinear spline in the average air pollution variable over all k lags, with $df_3$ degrees of freedom. As in previous discussions of nonlinear spline terms, there is no hard and fast rule for choosing $df_3$, but in subsequent discussion we have generally set it equal to 6 since this is large enough in most cases to show a clear nonlinearity but not so large that the model is distorted by evident overfitting.

The proposed model (3) is not as general as that of **[23]**, which will be considered in Section 3.3 of the present paper. The current model is simpler to fit and to interpret. On the other hand, by retaining the linear terms $\beta_2,\ldots,\beta_k$ in (2), the model is more general than that of Bell et al. [24], who used a spline term in the average ozone at lags 0 and 1, but without any "distributed lag" component.

The model formed by combining (1) and (3), based on lags 0,1,2, and 3, was fitted to all eight California air basins for both ozone and $PM_{2.5}$. Selected results are in Figures 11 and 12; results for other air basins are similar. In neither case is there any evidence of a systematic increase in risk with ozone or $PM_{2.5}$.

Figure 11. Nonlinear dependence of mortality on ozone for South Coast air basin. Blue dots: residuals from the model that includes long-term trends, day of week and meteorology, plotted against the air pollution variable (ozone). Red solid and dashed curves: implied change of relative risk with respect to ozone level 0.075 ppm (the current ozone standard), with pointwise 95% confidence bands.

Figure 12. Nonlinear dependence of mortality on $PM_{2.5}$ for San Francisco Bay air basin. Analogous to Figure 11, using the full meteorological model (including relative humidity), and a nonlinear model for the relationship between $PM_{2.5}$ and mortality.

## 3.3 Prediction, Methods and Results

With any large, complex, observational data set there are usually a lot of questions that can be considered. It is well-known that if you ask a lot of questions and employ a lot of modeling that false associations can result **[6, 7, 27]**. If effects are discovered, it is useful to know their relative importance. With all of this in mind we designed a factorial experiment to explore multiple questions and modeling that can be applied to the California data set.

The factorial experiment included, given in detail below: Health Endpoints, 4 levels; Air Basins, 8 levels; Air Quality variables including lags, 7 levels; Weather variables including lags, 9 levels; Time as Thin Plate Regression Splines. It is well-known that deaths follow a seasonal pattern, high in winter and low in summer. See **Figure 3**. Examination of the analysis results pointed to the importance of time in the analysis. All together over 78k models were fit. Our initial thinking was that if air quality was important for the health effects at issue, we should see a consistent pattern of results from the analysis across air basins and years.

The notation in this section differs somewhat from that in Section 3.2 so that the nature of the factorial modeling experiment is clear and that this section is somewhat self-contained. The basic data for Sections 3.1, 3.2 and 3.3 is the same.

In effect, the factorial experiment is a very large sensitivity study made possible by automated software using cloud computing.

**Methods**
**Spatial and Temporal Notation**

$$Y_{ijk} \sim Poisson(\mu_{ijk})$$

Let i = 1, 2, . . . , 8 indicate the following 8 air basins, Figure 1, in California:

Let j = 0, 1, . . . , 12 indicate the following 13 years, 2000, 2001, . . . , 2012.

Let k = 1, 2, . . . , $n_j$ indicate the $n_j$ days in year j, j = 0, 1, . . . , 12.

**Response and Covariates**
**Mortality, Response**
Using the death public use files provided by the state of California, we considered the total number of deaths in four different categories:

1. All cause deaths with accidents removed of individuals age 65 to 74, inclusive
2. All cause deaths with accidents removed for individuals age greater than or equal to 75
3. Death caused by diseases of the respiratory or circulatory systems individuals age 65 to 74, inclusive
4. Death caused by diseases of the respiratory or circulatory systems individuals age greater than or equal to 75

All subsequent methods were carried out for each of these four death outcomes. For the sake of describing the methods, let $Y$ generically indicate the response variable.

**Air Quality and Weather**

1. Daily Average PM$_{2.5}$ in micrograms per cubic meter denoted by $P_{ijk}$
2. Daily Max 8 Hour Overlapping Average Ozone in parts per billion denoted by $O_{ijk}$

**Prediction Models**

The following levels of response and covariates were considered in the subsequently defined generalized linear model:

**Air Quality**

1. Null
2. Ozone Current day $O_{ijk}$
3. Ozone Average of current day and lag-1 day $\dfrac{O_{ik} + O_{i,k-1}}{2}$
4. Ozone distributed lag nonlinear model: lag = 6, linear model in $O_{ijk}$ and B-Spline with 4 degrees of freedom in lag dimension denoted by $DLNM_{0-6}(O_{ijk})$ **[23]**
5. PM$_{2.5}$ Current day $P_{ijk}$
6. PM$_{2.5}$ Average of current day and lag-1 day $\dfrac{P_{ik} + P_{i,k-1}}{2}$
7. PM$_{2.5}$ Distributed lag nonlinear model: lag = 6, linear model in $P_{ijk}$ and B-Spline with 4 degrees of freedom in lag dimension $DLNM_{0-6}(P_{ijk})$

**Maximum Relative Humidity**

1. Null
2. Thin Plate Regression Spline (TPRS) **[24]** of current day $R_{ijk}$ denoted as $tprs(R_{ijk})$
3. $tprs(R_{ijk}) + tprs\left(\dfrac{R_{i,k-1} + R_{i,k-2} + R_{i,k-3}}{3}\right)$, where the second term is a TPRS of the average of lag-1,2, and 3 days

**Maximum Temperature**

1. Null
2. Thin Plate Regression Spline (TPRS) of current day $X_{ijk}$ denoted as $tprs(X_{ijk})$
3. $tprs(X_{ijk}) + tprs\left(\dfrac{X_{i,k-1} + X_{i,k-2} + X_{i,k-3}}{3}\right)$

**Minimum Temperature**

1. Null
2. Thin Plate Regression Spline (TPRS) of current day $Z_{ijk}$ denoted as $tprs(Z_{ijk})$
3. $tprs(Z_{ijk}) + tprs\left(\dfrac{Z_{i,k-1} + Z_{i,k-2} + Z_{i,k-3}}{3}\right)$

**Time**

$$tprs(k) + tprs(j)$$

A generalized linear model of the following form was fit for each of the levels of the respective response and covariates:

$$Y_{ijk} \sim Poisson(\mu_{ijk})$$
$$\log(\mu_{ijk}) = \beta_o + AirQuality_{ijk} + MaximumRelativeHumidty_{ijk} + MaximumTemperature_{ijk}$$
$$+ MinimumTemperature_{ijk} + tprs(k) + tprs(j)$$

Each year was individually held out of the model fitting procedure and the fitted model was to predict the response in the hold-out year. As a result, a total of $13 \times 4 \times 7 \times 3 \times 3 \times 3 \times 1 = 9828$ models were fit for each of the 8 air basins for a total of 78,624 models.

**Time-Series Analysis**
For each air basin and outcome the following two time series model were fit:

**Model 1: The Ozone + Weather Model**

$$Y_{ijk} \sim Poisson(\mu_{ijk})$$

$$\log(\mu_{ijk}) = \beta_o + DLNM_{0-6}(O_{ijk}) + tprs(R_{ijk}) + tprs\left(\frac{R_{i,k-1} + R_{i,k-2} + R_{i,k-3}}{3}\right)$$

$$+ tprs(X_{ijk}) + tprs\left(\frac{X_{i,k-1} + X_{i,k-2} + X_{i,k-3}}{3}\right)$$

$$+ tprs(Z_{ijk}) + tprs\left(\frac{Z_{i,k-1} + Z_{i,k-2} + Z_{i,k-3}}{3}\right)$$

$$+ tprs(k) + tprs(j)$$

**Model 2: The PM$_{2.5}$ + Weather Model**

$$Y_{ijk} \sim Poisson(\mu_{ijk})$$

$$\log(\mu_{ijk}) = \beta_o + DLNM_{0-6}(P_{ijk}) + tprs(R_{ijk}) + tprs\left(\frac{R_{i,k-1} + R_{i,k-2} + R_{i,k-3}}{3}\right)$$

$$+ tprs(X_{ijk}) + tprs\left(\frac{X_{i,k-1} + X_{i,k-2} + X_{i,k-3}}{3}\right)$$

$$+ tprs(Z_{ijk}) + tprs\left(\frac{Z_{i,k-1} + Z_{i,k-2} + Z_{i,k-3}}{3}\right)$$

$$+ tprs(k) + tprs(j)$$

For each combination of basin and response, the cross reduce function in the DLNM package in R was used to obtain the overall cumulative effect of ozone and the overall cumulative effect of PM$_{2.5}$. The overall cumulative ozone effect is the risk of death at a particular value of ozone relative to an ozone level of 50 ppb. The overall cumulative PM$_{2.5}$ effect is the risk of death at a particular value of PM$_{2.5}$ relative to a PM$_{2.5}$ level of 20 micrograms per cubic meter. The coefficient and variance of the overall cumulative effect for each combination of basin, response and air quality variable will be obtained.

For each response and air quality variable, the 8 basin specific cumulative effect coefficients were combined using a meta-analysis framework as described in Gasparrini [25]. Let $\theta_i$ be the basin specific cumulative effect and $S_i$ be the corresponding variance of the cumulative effect for each basin. We assume the following model for each basin

$$\theta_i \mid \theta, \sigma^2, S_i \sim N(\theta, S_i + \sigma^2),$$

where $\theta$ is the pooled effect estimate and $\sigma^2$ is interpreted as the between basin variance. We use restricted maximum likelihood to estimate the parameters $\theta$ and $\sigma^2$.

**Results: Prediction Models**

Note due to missing covariate data in the calendar year 2000 results for that year as the hold-out year are omitted due to large numbers of missing values for predictions.

By partitioning the air quality variable into two groups, Ozone (levels 1,2,3, and 4) and $PM_{2.5}$ (levels 1,5,6, and 7), 108 models were isolated for each combination of air quality group, basin, year, and response (4 levels of air quality X 3 levels of Max. Temp. X 3 levels of Min. Temp. X 3 levels of Rel. Hum. = 108 models). Note 27 models appear in both groups because of the null level (level 1) of the air quality variable.

First, the observed values for each combination of basin, year, and response were plotted (open circles) and the predictions from the 108 models were added to the same plot (solid red lines). Consider the results for the number of deaths caused by diseases of the respiratory or circulatory systems individuals age greater than or equal to 75 for the South Coast air basin for the Ozone group, **Figure 13.**

Figure 13. Model hold out predictions for each year except 2000. "o" are the observed deaths and the red overlay are the model predictions. The models do not explain much of the variability and are essentially identical to one another.

Despite the various forms of the 108 models, the variability of the predicted values is relatively small as illustrated by the overlapping red lines. Because the predictions are point estimates, prediction intervals that account for uncertainty would overlap and thus make the predictions virtually indistinguishable **[27].** That is, in terms of predictive performance, the models perform similarly. Note a similar result for the other air basins in both ozone groups and the $PM_{2.5}$ groups regardless of outcome **(Sup Figures A1-A32 and B1-B32).**

Second, the mean squared prediction error (MSPE) was obtained for each model using the data from the year that was held out. For each combination of air quality group, basin, year, and response, the MSPE of the model that only includes time as a covariate, $MSPE_t$, was used to calculate the ratio

$$\frac{MSPE_m}{MSPE_t},$$

for each value $m = 1, \ldots, 108$ indexing the 108 models considered for that combination of air quality group, basin, year and response. For a given model, if the ratio is greater than 1 then the model that only included time had a smaller MSPE and if the ratio is less than 1 then the corresponding model had an MSPE smaller than the model that only included time. A boxplot of the 108 MSPE ratios for each combination of air quality group, basin, year, and response are presented **(Figure 14** and **Sup 3.3 C and Sup 3.3D, 32 scatter graphs each.)**

Figure 14. Box plots of hold one year out of mean square prediction errors, MSPE. The predictions are made by varying the modeling variables.

Consider the ratios of the MSPE of each of the 108 models for the same subset of data, number of deaths caused by diseases of the respiratory or circulatory systems individuals age greater than or equal to 75 for the South Coast air basin for Ozone group, **Figure 13.** Recall a value greater than 1 indicates the model had an MSPE larger than the model that include time effects only and if the value is less than 1 then the model had an MSPE smaller than the model that included time only. Note that in general the ratio of the MSPE relative to the MSPE of the model with time only fell between 0.98 and 1.02. Moreover, the variability of the ratio changes depending on which year is held out. In addition, the form model with the best MSPE (i.e. the smallest ratio) was not the consistent across year **(Supplementary File 03 Prediction analysis results).** In summary, the boxplots indicate that the differences in point-estimate predictions for hold-out years are small and there is not a consistent best form of the model. This result is consistent across response variable, air quality group, and basin **(Supplementary Figures C1-C32, D1-D32, and Supplementary File 1).**

**Figures 15 and 16** give the estimate dose response over the eight air basins. The solid blue lines gives composite slopes that the confident limits for the composite slopes overlap a slope of 1.0, no effect.

Figure 15. Ozone. The time-series analysis was conducted to obtain an estimated relationship between the air quality variables and the response variables in each basin. The basin specific estimates were combined in a meta-analysis framework to obtain the pooled cumulative effect. The relationships between the response variables and ozone are plotted. The first-stage basin-specific relationships are represented by the dashed lines. The pooled estimates with 95% confidence intervals are represented by the solid blue lines and shaded gray region, respectively.

Figure 16 $PM_{2.5}$. The time-series analysis was conducted to obtain an estimated relationship between the air quality variables and the response variables in each basin. The basin specific estimates were combined in a meta-analysis framework to obtain the pooled cumulative effect. The relationships between the response variables and ozone are plotted. The first-stage basin-specific relationships are represented by the dashed lines. The pooled estimates with 95% confidence intervals are represented by the solid blue lines and shaded gray region, respectively.

## 4.0 Discussion

In this paper we analyze daily death data for the eight most populous air basins in California for associations with air quality. We found no associations using our moving median analysis method and no consistent associations with our regression-based time series analysis. Moreover, we assessed the predictive capability of various models using a leave-one-year-out cross validation strategy. We found air quality variables most often do not add to the predictive ability of the model. Even when the predictive ability is improved, the improvement is negligible relative to a model that only uses time of year. Also the form of the air quality variable that improves prediction is inconsistent across basin/year combinations. In short, we were unable to find a consistent and meaningful relationship between air quality and acute death in any of the eight California air basins considered.

The time series methods used in Sections 3.2 and 3.3 have become standard in studies of this nature. Daily death counts are fitted using a generalized linear model with log link and either Poisson or "quasipoisson" distributional form. The latter assumes the same mean-variance structure as the Poisson distribution but with one additional parameter representing overdispersion. In the analyses of this paper, the overdispersion parameter is typically of the order of 1.05 to 1.07, indicating only a slight departure from the Poisson distribution. However, the standard errors are slightly larger when the overdispersion parameter is included and this is therefore recommended as a conservative approach.

The covariates included in the analysis are long-term trend, day of week and three meteorological variables: daily maximum temperature, daily minimum temperature and relative humidity. The well-known NMMAPS dataset used daily mean temperature and dewpoint as the two meteorological variables, but in other respects the analysis is the same. In particular, for each meteorological variable we have included the current day's value and the average of the three previous days' values, each modeled nonlinearly through a smoothing spline. The degrees of freedom of these smoothing splines, as well as the one involving long-term trend, have been varied to allow us to study the sensitivity to that parameter.

The trickiest aspect of the analysis is in deciding which lag or lags of the air pollution variable to include. Unlike the choice of degrees of freedom in the smoothing spline, the choice of lags for the air pollution analysis does appear to have a substantial effect on the estimated coefficients. In these analyses, we have tried a total of ten different combinations of lags for the single-air basin analyses, and then used the results of that to guide the choice of lags for the combined analyses. Trying a wide variety of different lag structures and only using the one that gives the largest coefficient or the most statistically significant result has a flavor of "data snooping" and could lead to biased results. We have tried to mitigate that effect by only considering results for which the p-value is well under the standard 0.05, but there are not many such cases where even this mild criterion is satisfied.

For ozone, we were unable to find any significant result for the South Coast air basin, and only for San Francisco Bay when the humidity variable was excluded, and with a relatively mild p-value (0.018). When the results are combined across all air basins, there is no effect of ozone on mortality.

For PM2.5, there were several statistically significant results in South Coast but with negative coefficients, which does not make sense biologically and could be a model selection artifact. In San Francisco Bay, there were some statistically significant results with a positive coefficient, but only for models in which relative humidity was excluded. Given the inconsistent results and the relatively mild p-values associated with them, it seems likely that these results are a case of spurious statistical significance. When combined across all air basins, there is no statistically significant effect for PM2.5.

To establish a direct comparison with the analyses of the NMMAPS dataset in Smith **[9],** the results of that paper have been re-derived using the exact same statistical method used in the present paper (in particular, the same meteorological variables). Consistent with the results of the present paper, none of the individual-city results for California using the NMMAPS dataset showed a statistically significant effect, nor did the combined result of all California cities. However, the present method of analysis still shows a statistically significant "national" effect for ozone when applied to the whole NMMAPS dataset, although smaller than that reported in Smith **[9]**. The fact that the national estimate is smaller using tmax, tmin and relative humidity as the meteorological variables could imply that these variables do a better job than the original NMMAPS variables (daily mean temperature and dewpoint) of capturing the confounding effect of ozone with meteorology. Moreover, Smith **[9]** noted spatial variation of the ozone-mortality coefficient across cities, possible explanatory variables being the percentage of residences with central air conditioning (high in California) and the use of public transportation (high use of public transportation could correspond to high exposure to ambient pollutants, but use of public transportation is generally low in California). Because of this spatial variation, Smith **[9]** questioned whether it made sense to compute a "national" estimate in the face of clear evidence that the effect is not, in fact, national in scope. The results of the present paper add to the previous results by confirming that California data for 2000-2012, most of which lies after the end of the NMMAPS data period, still does not show any statistically significant relationship between mortality and either ozone or PM$_{2.5}$.

The analyses of Section 3.3 extend these results by considering a still wider range of models and also assessing predictive power of the models in a cross-validation context. For most analyses, including the air pollution variable does not improve the predictive ability of the model.

It is worth considering how these results related to other studies of ozone and especially PM$_{2.5}$, especially those regarding long-term chronic effects. There is contradictory literature on chronic effects of air pollution on deaths for the entire US. Eight papers were cited in [28], references 4–11 in their paper, saying, "Associations between long-term exposure to fine particulate air pollution and mortality have been observed ... more recently, in cohort-based studies.... all support the view that relatively prompt and sustained health benefits are derived from improved air quality". On the other hand, Enstrom **[29],** after citing papers supporting an association says, "Other cohort studies have also examined mortality associations with PM2.5 and other pollutants ... with somewhat different findings." Enstrom cites four papers that cast doubt on the claim. Enstrom has a particular interest in California and he extracted risk ratio summary data for all cause deaths for California from a number of papers **[30]**. See **Table 9**. The average risk ratio was 0.9979 with a standard error of 0.0126 for California.

|  | Years | Risk Ratio | Confidence Limits |
|---|---|---|---|
|  |  |  |  |
| McDonnell 2000 | 1976-1992 | 1.03 | 0.95 -1.12 |
| Krewski 2000 | 1982-1989 | 0.872 | 0.805-0.944 |
| Enstrom 2005 | 1973-1982 | 1.039 | 1.010-1.069 |
| Enstrom 2005 | 1983-2002 | 0.997 | 0.978-1.016 |
| Enstrom 2006 | 1973-1982 | 1.061 | 1.017-1.106 |
| Enstrom 2006 | 1983-2002 | 0.995 | 0.968-1.024 |
| Zeger 2008 | 2000-2005 | 0.989 | 0.970-1.008 |
| Jerrett 2010 | 1982-2000 | 0.994 | 0.965-1.025 |
| Krewski 2010 | 1982-2000 | 0.96 | 0.920-1.002 |
| Krewski 2010 | 1982-2000 | 0.968 | 0.916-1.022 |
| Jerrett 2011 | 1982-2000 | 0.994 | 0.965-1.024 |
| Jerrett 2011 | 1982-2000 | 1.002 | 0.992-1.012 |
| Lipsett 2011 | 2000-2005 | 1.01 | 0.95 -1.09 |
| Ostro 2011 | 2002-2007 | 1.06 | 0.96 -1.16 |

Table 9. All Cause risk ratios for PM$_{2.5}$ deaths in California **[30].**

Chay et al. **[31]** looked at a reduction in air pollution due to the Clean Air Act, focusing on total suspended particulates (TSPs). Counties out of compliance were given stricter air pollution reduction goals. This action by the EPA created a so called natural experiment **[32].** The EPA selected counties that they judged needed to reduce air pollution levels. Air pollution levels were reduced in these counties, but there was no reduction in deaths after adjustments for covariates. They concluded "…regulatory status is associated with large reductions in TSPs pollution but has little association with reductions in either adult or elderly mortality." Another paper **[33]** also found that a reduction in PM$_{2.5}$ does not lead to a reduction in deaths. No association of PM$_{2.5}$ with longevity in western US was claimed in **[34].** Many others have found no association of chronic deaths with PM2.5 in California. See **Table 9** and the references cited therein.

Another analysis that tried to reconcile claims related to acute and long-term effects was Greven et al. **[35]**. They developed a Poisson regression model to estimate two regression coefficients, a "global" coefficient that measures the association between national trends in pollution and mortality, and a "local" coefficient, derived from space by time variation, that measures the association between location-specific trends in pollution and mortality adjusted by the national trends. In their results, they find strong statistically significant evidence for the global coefficient, but they acknowledge that this could be explained by a variety of confounding factors operating at a national level. In contrast, the local coefficient, being based on correlating spatial variations in mortality with corresponding spatial variations in $PM_{2.5}$, is supposed to be largely free of confounding variables. However, they did not find statistical significance for the local regression coefficient. This result therefore raises the question of whether there is any association between long-term mortality and $PM_{2.5}$ that cannot be attributed to confounding variables. An earlier analysis in **[36]** raised similar concerns. In summary, a number of papers that take covariates into account **[31, 33, 35-37]** lead to the conclusion that there is no association of air quality with deaths that could not be explained by confounding variables.

Many authors have noted "geographic heterogeneity", that the measured effect of air quality is not the same in different locations. There is overwhelming evidence for the existence of geographic heterogeneity **[8, 30, 31, 33, 38]**. Multiple authors **[9, 34, 37, 40]** have not found any association of air quality with acute deaths in California. Enstrom **[29]** found no association with chronic deaths in California. In the absence of an adequate explanation of why such geographical heterogeneity occurs, it is hard to make the case for a simple causal relationship between air quality and deaths. Indeed, Greven et al. **[35]** suggest that differences in locations (geographic heterogeneity) are most likely due to differences in covariates that are not connected with air pollution such as age distributions, income and smoking.

Another recent paper by Milojevic et al. **[41]** effectively removes heart attacks and stroke as a possible etiology for acute air quality deaths. This paper uses a very large UK data set. They track heart attacks and strokes essentially to the hour, but they find no association of heart attacks with ozone or $PM_{2.5}$. We also note that cardiovascular notations on a death certificate are not considered reliable, for example **[42],** which reports that about half of all heart attacks are missed and heart attacks are falsely diagnosed about 25% of time, hence our primary focus on all-cause (non-accidental) deaths. The cardiovascular diagnosis is largely a matter of convenience as the person filling out the certificate is required to put something down.

In summary, although EPA publications such as **[43]** highlight the large number of papers supporting a harmful association between particle pollution and public health, in fact, the evidence is not so clear cut: we have highlighted numerous studies where variations in mortality could be better explained by covariates not related to air pollution, or by simple geographic heterogeneity, without implying a causal association with air pollution. The present paper adds to that literature by showing that, even by applying a variety of different statistical methods, we were unable to find any association between daily mortality and either $PM_{2.5}$ or ozone in what we believe to be the only public dataset based entirely on data since the 1997 revision of air pollution standards. Therefore, we conclude that the case for further revision of those standards is unproven at the present time.

**03 Figures for Air quality and acute deaths in California**

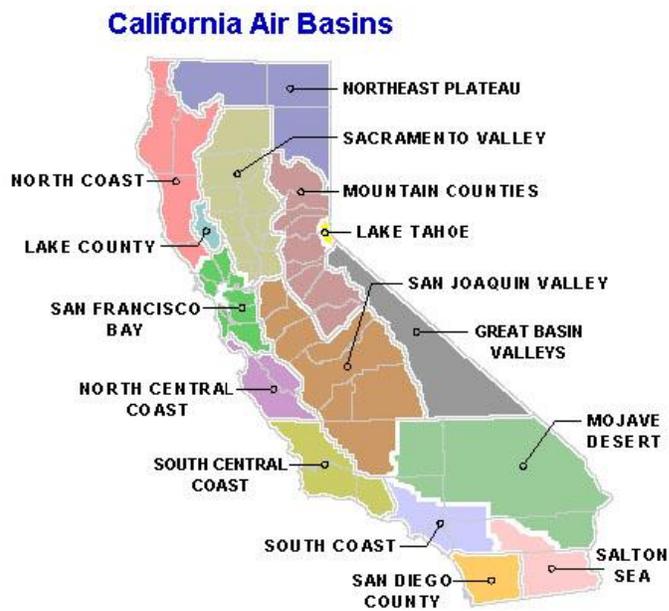

**Figure 1:** Map of California Air Basins (Source: Webpage of the California Air Resources Board)

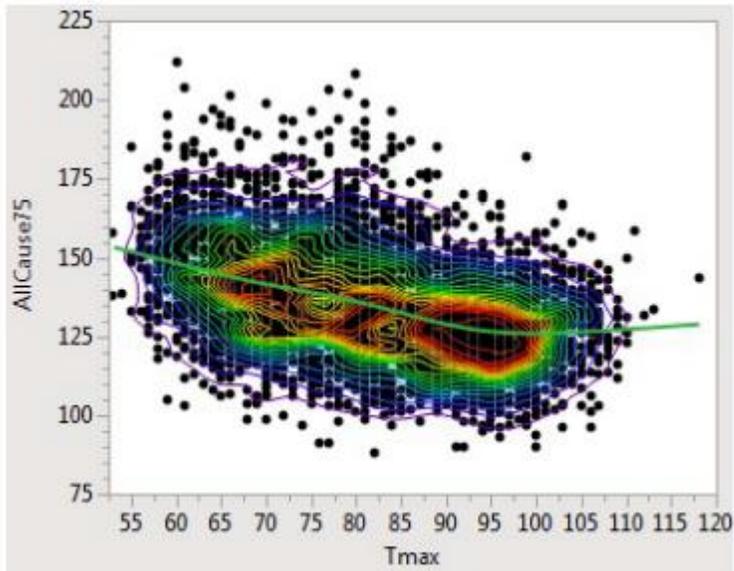

Figure 2. All Cause daily deaths of people 75 and older are plotted against the maximum daily temperature. Each dot represents the number of daily deaths. Accidental deaths are removed. Non-parametric density contours are plotted to help overcome the overprinting.

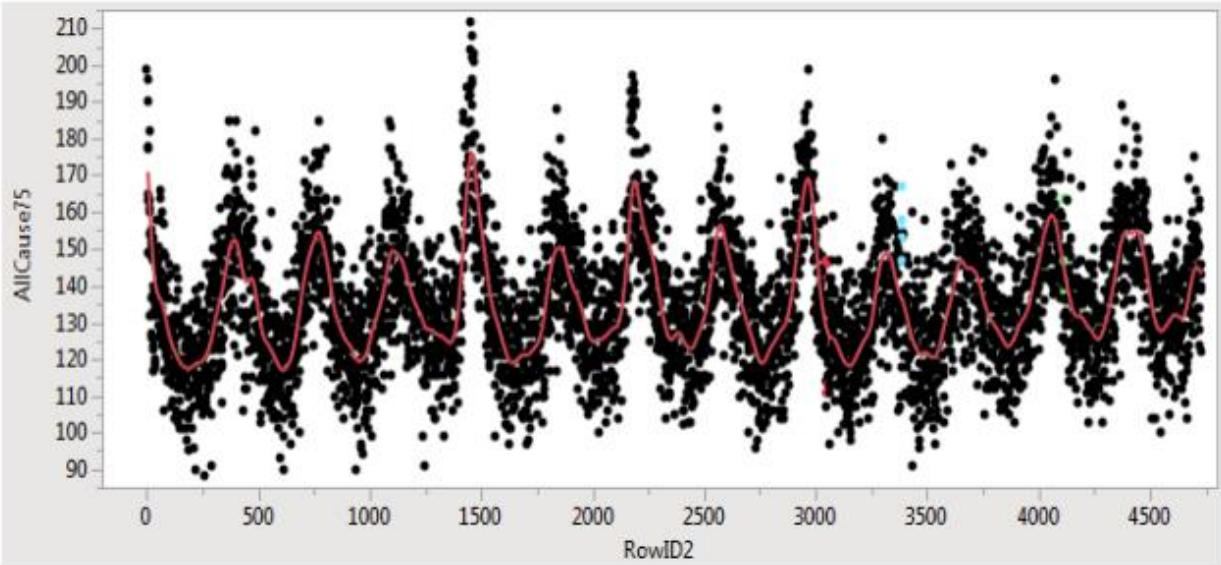

Figure 3. Daily all cause deaths are plotted against time, 2000-2012. A spline is fit to the data to help visually track the density of the points.

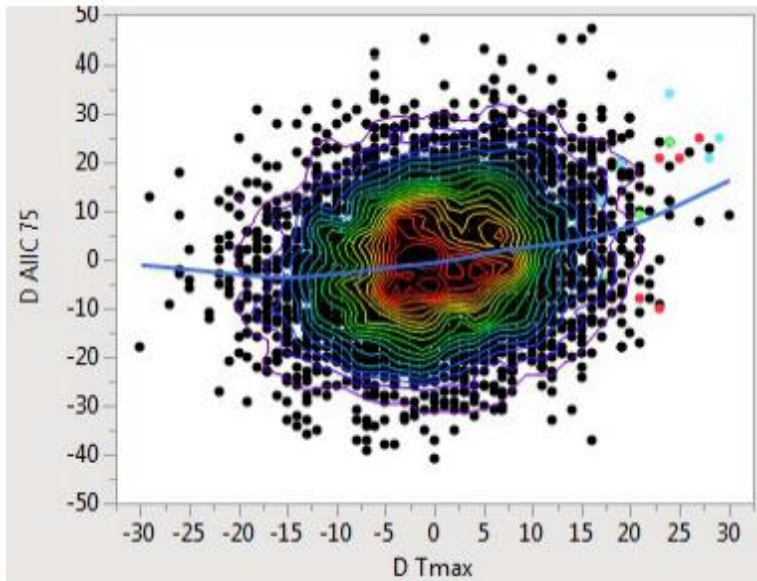

Figure 4. Deviations of daily deaths are plotted against deviation of daily temperature. A spline indicates that as the there is a spike in daily maximum temperature there is an increase in daily deaths. Three sets of points are colored, red, green and aqua where the spike in temperature is 15-28 degrees higher than the moving median.

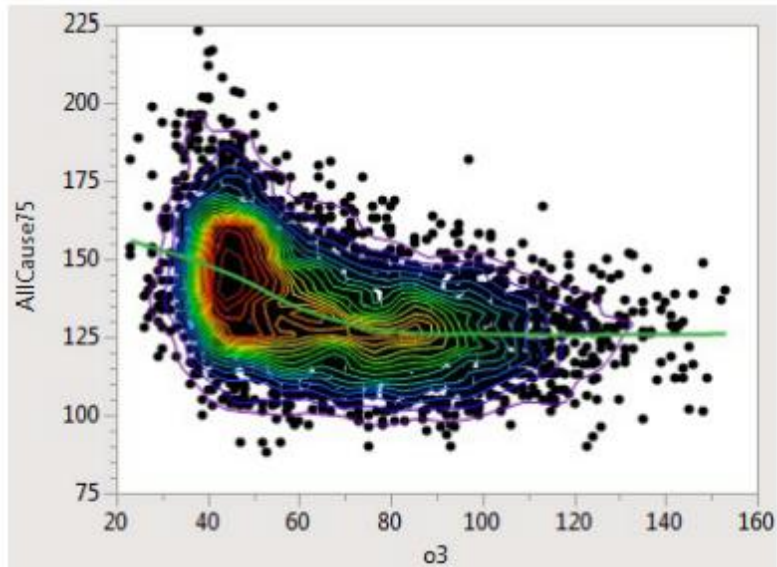

Figure 5. All Cause daily deaths of people 75 and older are plotted against the maximum 8-hour daily ozone. Each dot represents the number of daily deaths. Accidental deaths are removed. Non-parametric density contours are plotted to help overcome the overprinting.

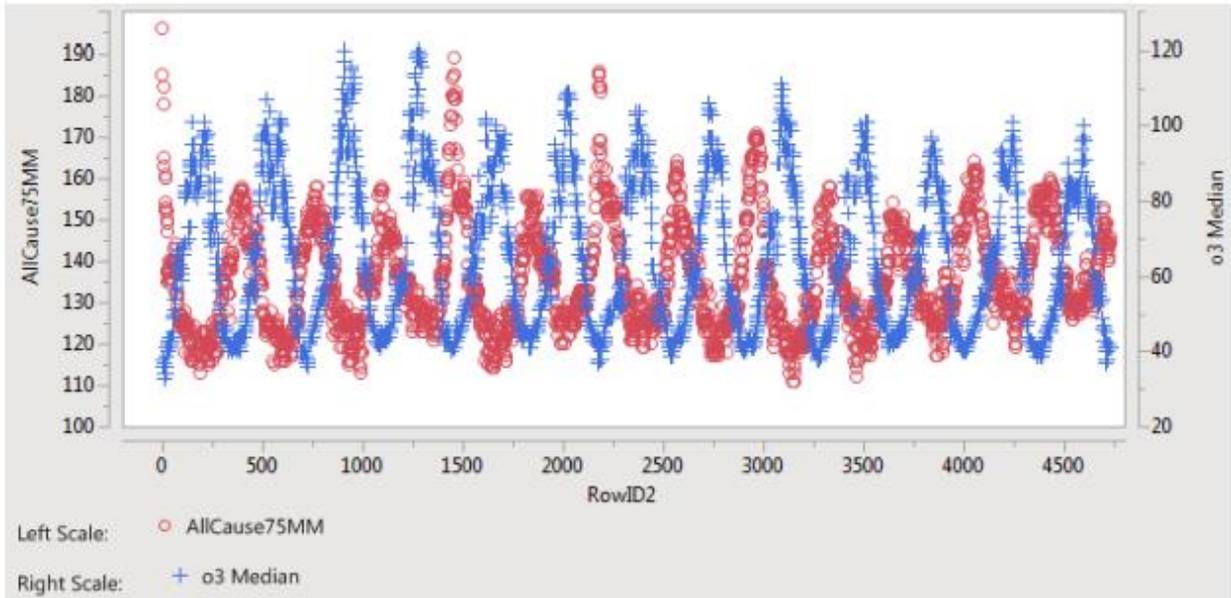

Figure 6. The moving medians for all cause deaths and ozone are plotted against time, 13 years, red o for deaths and blue + for ozone.

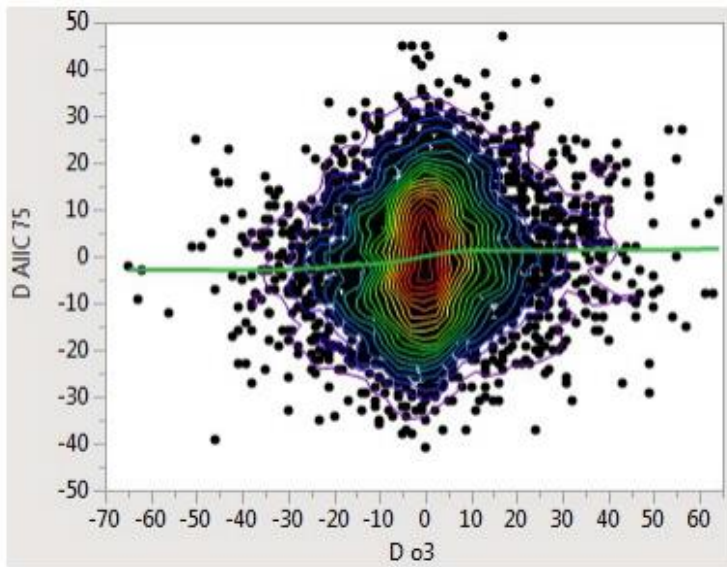

Figure 7. Deviations of daily deaths are plotted against deviation of daily ozone. A spline indicates that when the there is a spike in daily ozone, there is little or no effect on daily deaths. The density does not go from lower left to upper right. Dramatic increases in ozone have essentially no effect on deaths.

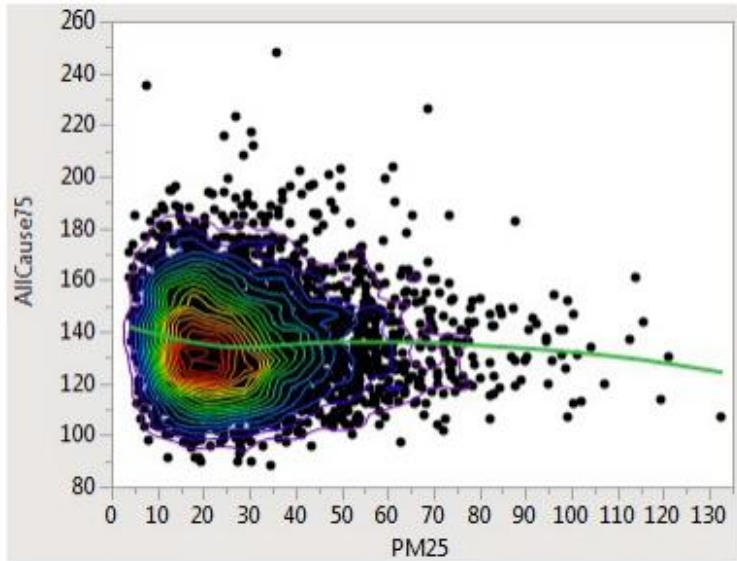

Figure 8. All Cause daily deaths of people 75 and older are plotted against the PM2.5. Each dot represents the number of daily deaths. Accidental deaths are removed. Non-parametric density contours are plotted to help overcome the overprinting. A spline fit to the data indicates no association of deaths with PM2.5 levels.

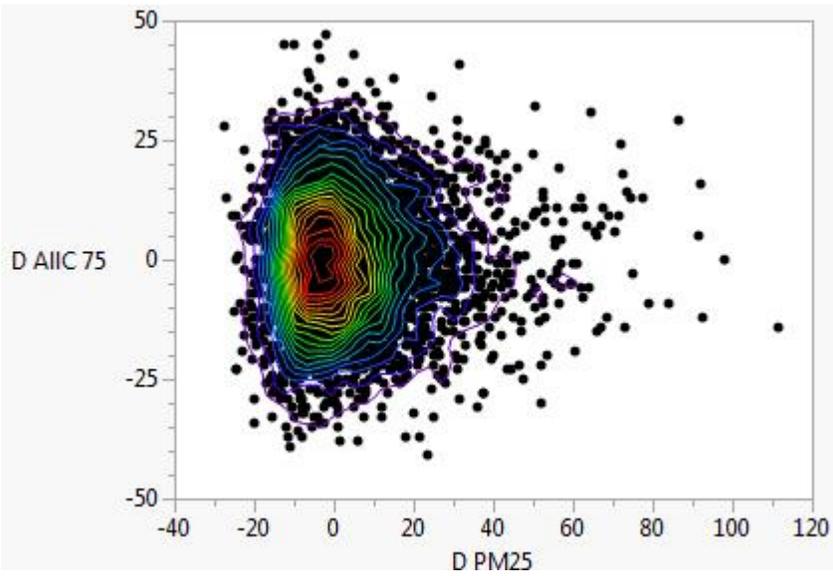

Figure 9. Deviations of All Cause daily deaths of people 75 and older are plotted against the deviations of PM2.5 from yearly medians. There is no indication of any association.

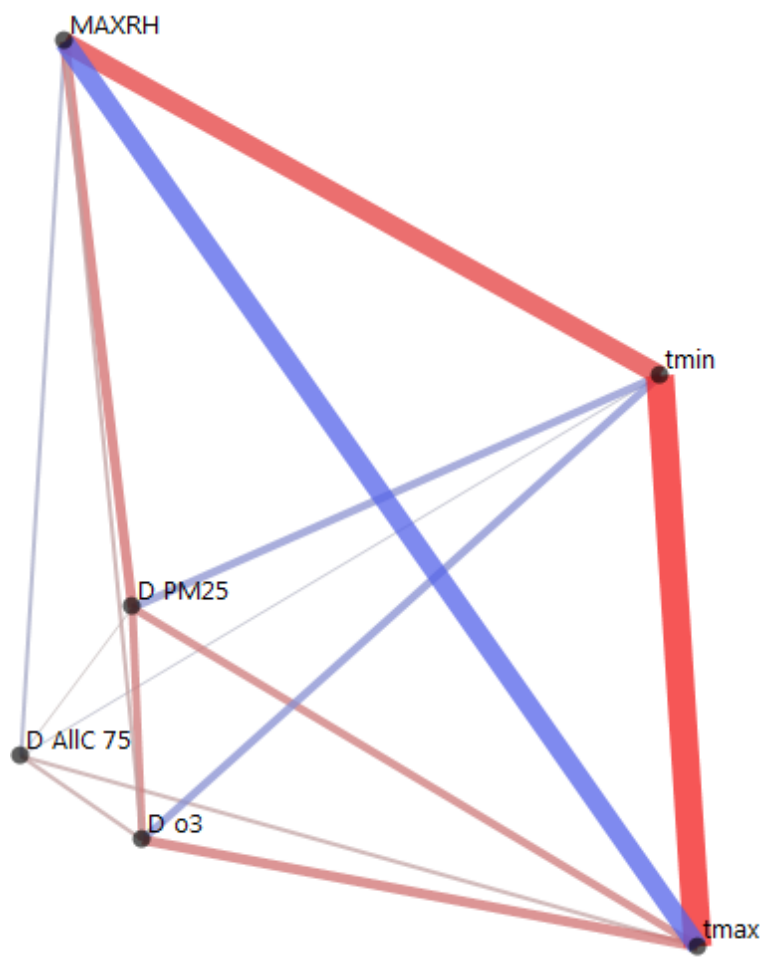

Figure 10. Partial correlation diagram among the moving median deviations. Partial correlations between all cause mortality and air pollution variables are inconsequential.

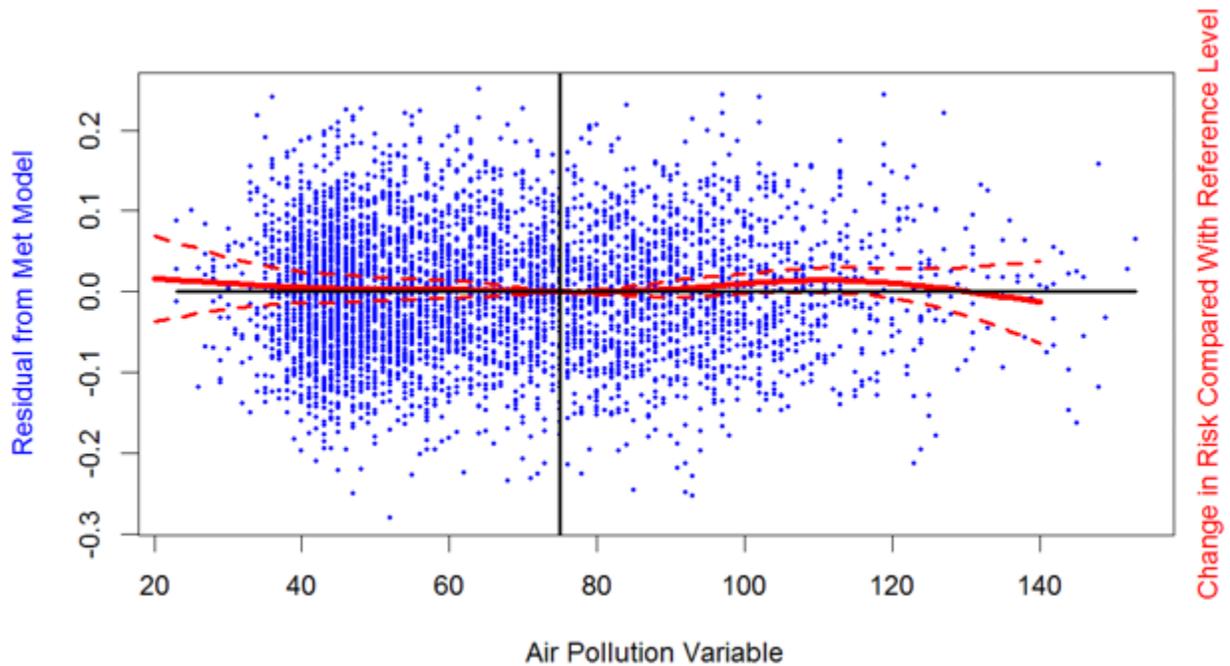

Figure 11. Nonlinear dependence of mortality on ozone for South Coast air basin. Blue dots: residuals from the model that includes long-term trends, day of week and meteorology, plotted against the air pollution variable (ozone). Red solid and dashed curves: implied change of relative risk with respect to ozone level 0.075 ppm (the current ozone standard), with pointwise 95% confidence bands.

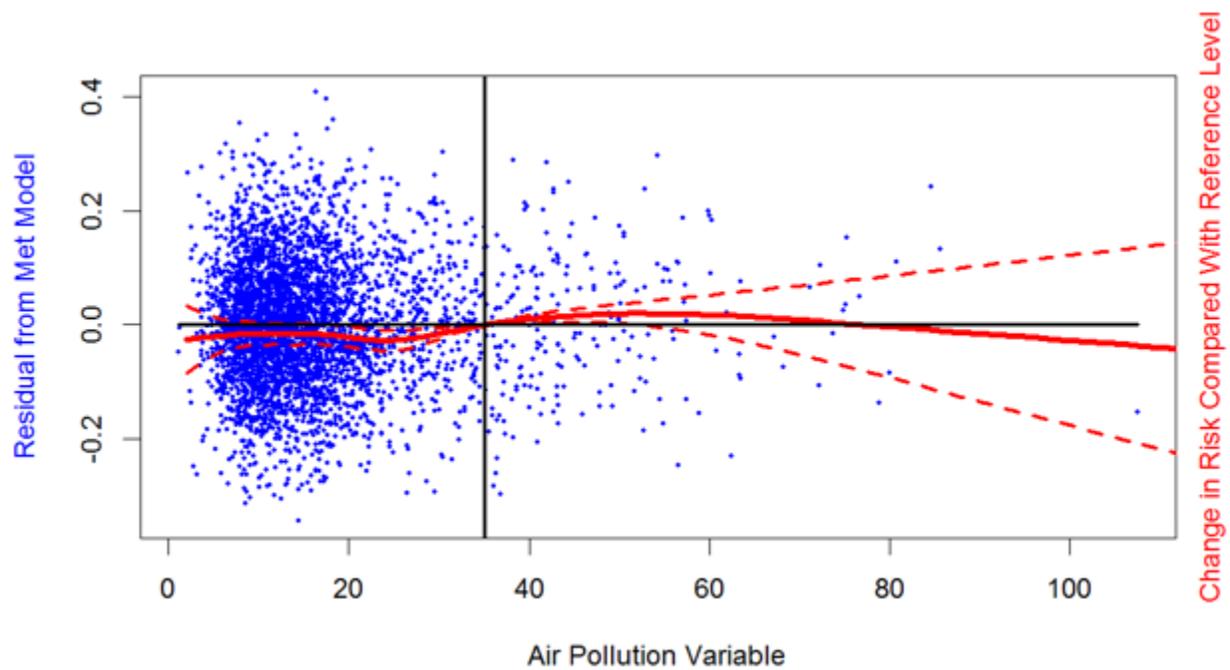

Figure 12. Nonlinear dependence of mortality on $PM_{2.5}$ for San Francisco Bay air basin. Analogous to Figure 11, using the full meteorological model (including relative humidity), and a nonlinear model for the relationship between $PM_{2.5}$ and mortality.

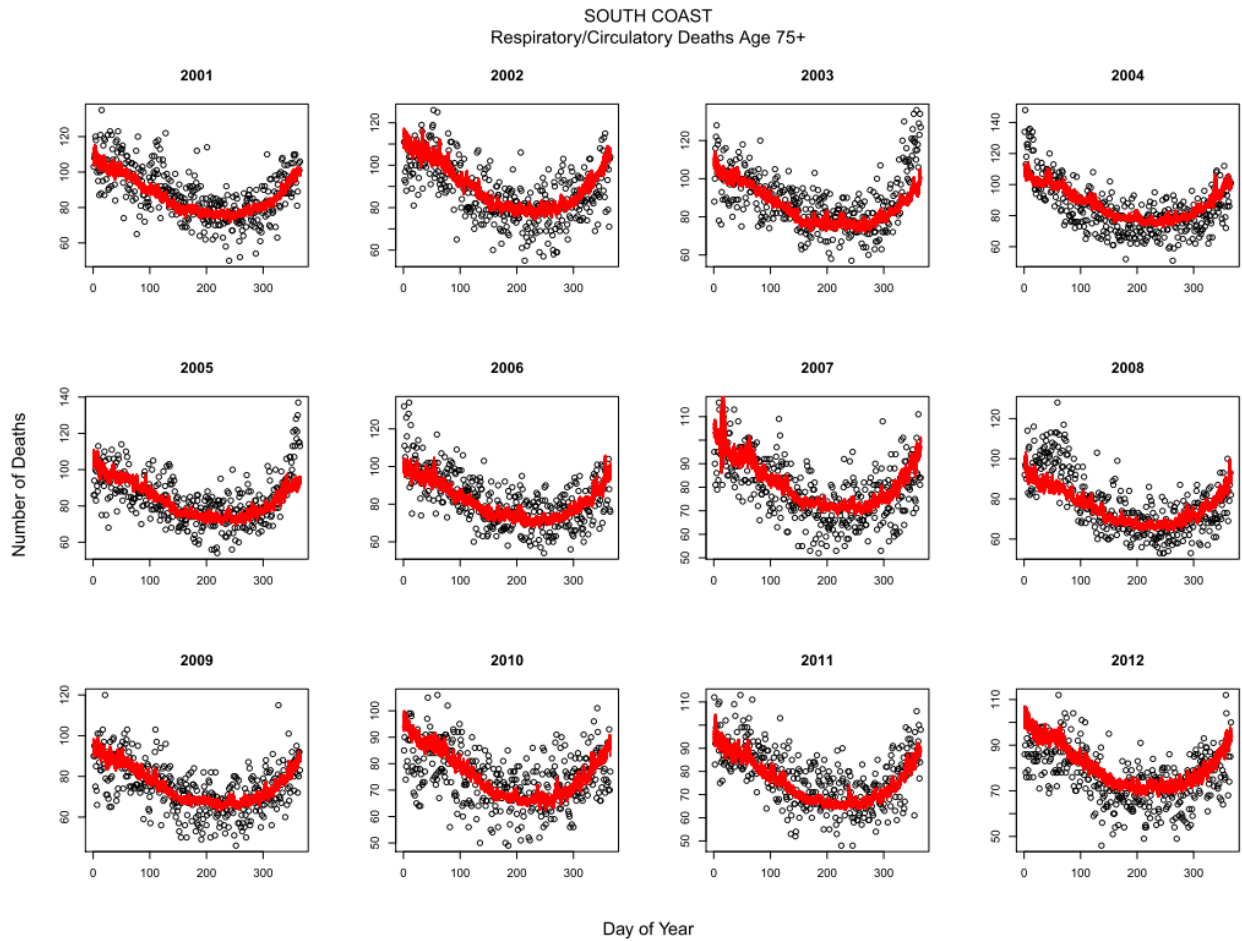

Figure 13. Model hold out predictions for each year except 2000. "o" are the observed deaths and the red overlay are the model predictions. The models do not explain a lot of the variability and are essentially identical to one another.

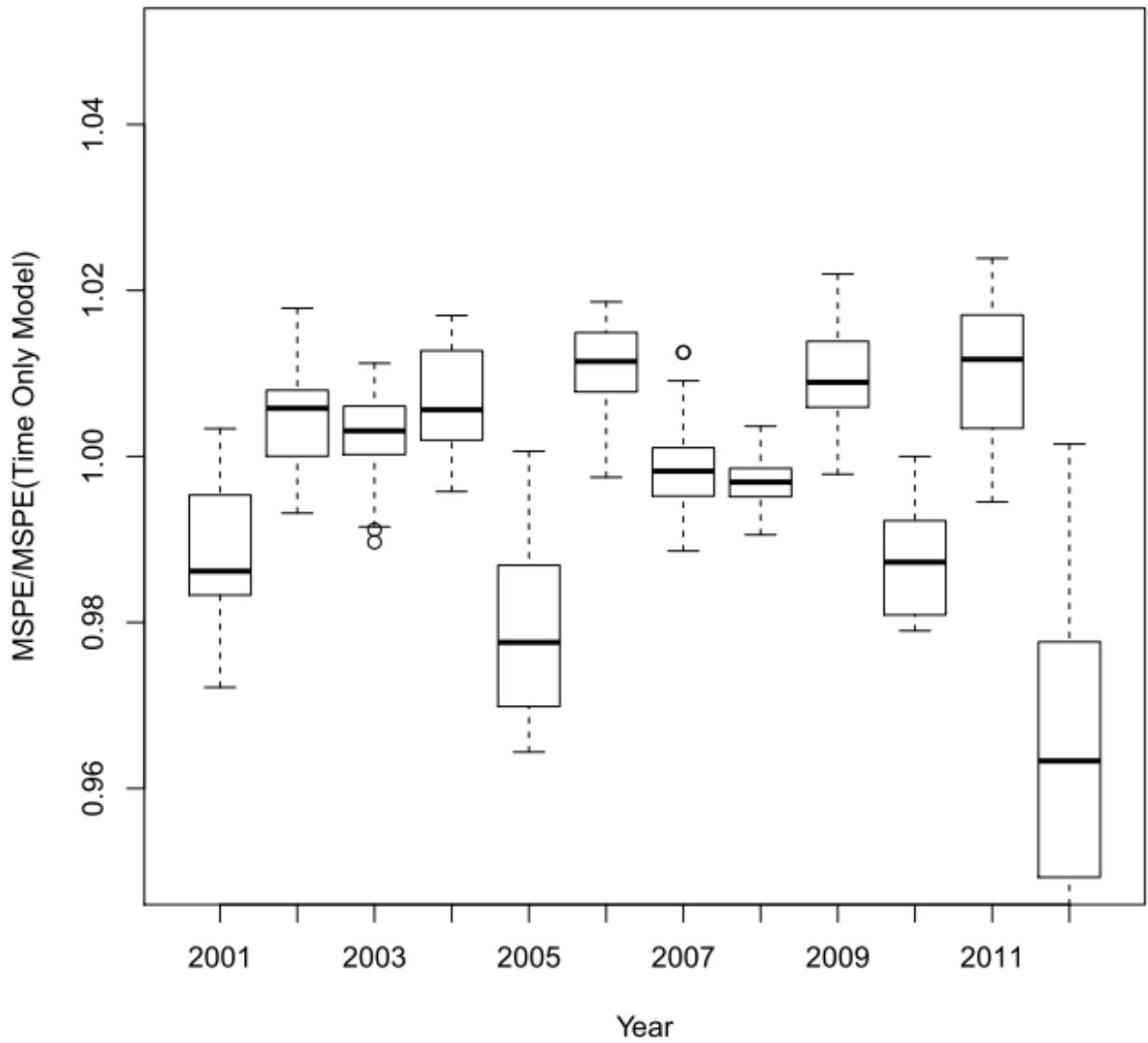

Figure 14. Box plots of hold one year out of mean square prediction errors, MSPE. The predictions are made by varying the modeling variables.

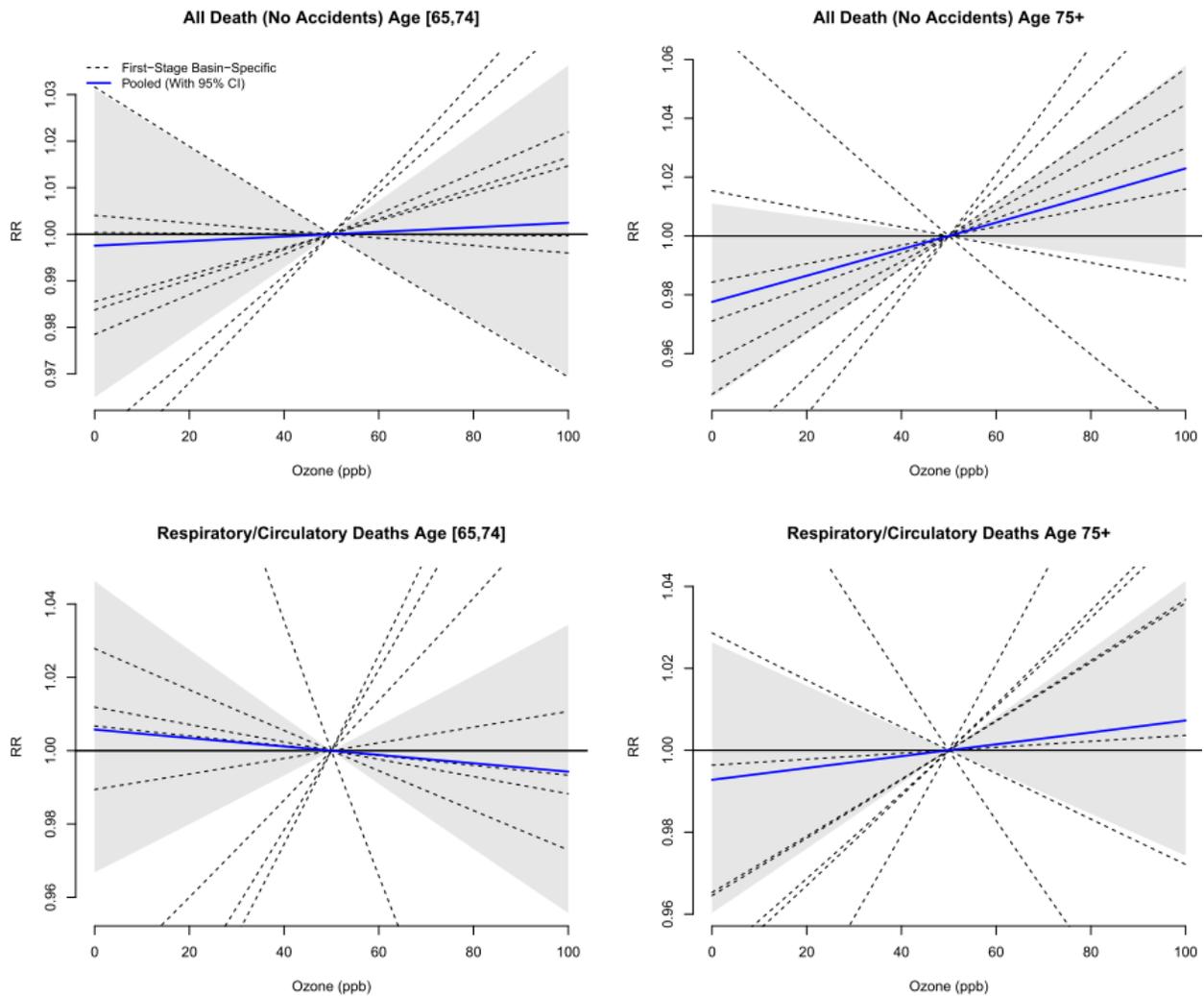

Figure 15. Ozone. The time-series analysis was conducted to obtain an estimated relationship between the air quality variables and the response variables in each basin. The basin specific estimates were combined in a meta-analysis framework to obtain the pooled cumulative effect. The relationships between the response variables and ozone are plotted. The first-stage basin-specific relationships are represented by the dashed lines. The pooled estimates with 95% confidence intervals are represented by the solid blue lines and shaded gray region, respectively.

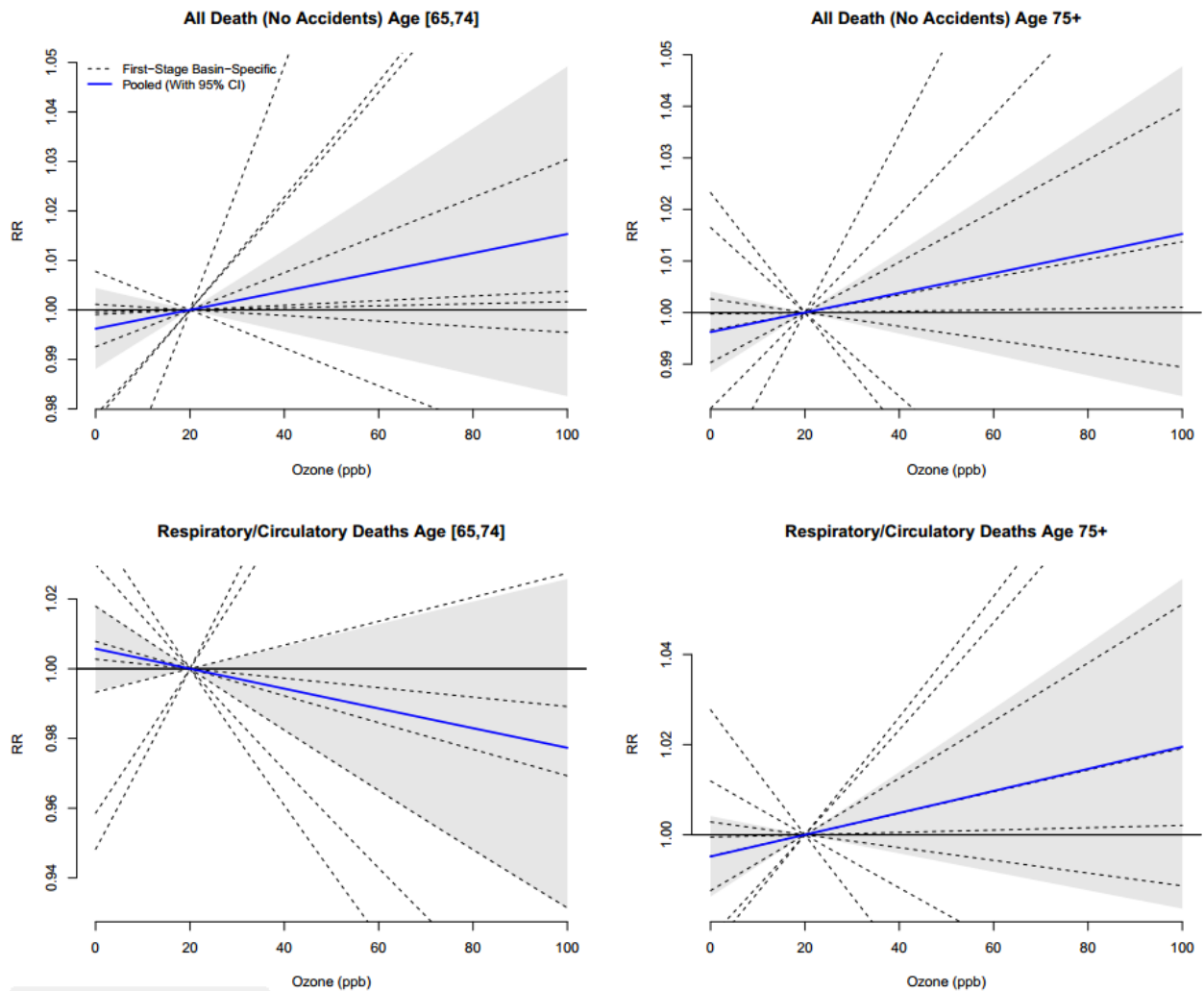

Figure 16. PM$_{2.5}$. The time-series analysis was conducted to obtain an estimated relationship between the air quality variables and the response variables in each basin. The basin specific estimates were combined in a meta-analysis framework to obtain the pooled cumulative effect. The relationships between the response variables and ozone are plotted. The first-stage basin-specific relationships are represented by the dashed lines. The pooled estimates with 95% confidence intervals are represented by the solid blue lines and shaded gray region, respectively.